\documentclass[aps,showpacs,tightenlines,twocolumn,nofootinbib,nobibnotes,superscriptaddress]{revtex4-1}
\usepackage{amsmath,amssymb,amsfonts,bm}
\usepackage{graphicx}
\usepackage{epstopdf}
\usepackage{dcolumn}
\usepackage{mathrsfs}
\usepackage{tgtermes}
\usepackage[colorlinks=true,linkcolor=blue,citecolor=blue, urlcolor=blue,bookmarks=false]{hyperref}
\usepackage{lipsum}
\usepackage[marginal]{footmisc}

\bibpunct{[}{]}{,}{n}{}{}
\usepackage{multirow}
\begin{document}
\title{Structural disorder-induced topological phase transitions in quasicrystals}
\date{\today}
\author{Tan Peng}
\altaffiliation{These authors contributed equally to this work.}
\affiliation{Key Laboratory of Artificial Micro- and Nano-Structures of Ministry of Education, School of Physics and Technology, Wuhan University, Wuhan 430072, China}

\affiliation{Hubei Key Laboratory of Energy Storage and Power Battery, and School of Mathematics, Physics and Optoelectronic Engineering, Hubei University of Automotive Technology, Shiyan 442002, China}

\author{Yong-Chen Xiong}
\altaffiliation{These authors contributed equally to this work.}
\affiliation{Hubei Key Laboratory of Energy Storage and Power Battery, and School of Mathematics, Physics and Optoelectronic Engineering, Hubei University of Automotive Technology, Shiyan 442002, China}

\author{Chun-Bo Hua}
\affiliation{School of Electronic and Information Engineering, Hubei University of Science and Technology, Xianning 437100, China}

\author{Zheng-Rong Liu}
\affiliation{Department of Physics, Hubei University, Wuhan 430062, China}

\author{Xiaolu Zhu}
\affiliation{Key Laboratory of Artificial Micro- and Nano-Structures of Ministry of Education, School of Physics and Technology, Wuhan University, Wuhan 430072, China}

\author{Wei Cao}
\affiliation{Key Laboratory of Artificial Micro- and Nano-Structures of Ministry of Education, School of Physics and Technology, Wuhan University, Wuhan 430072, China}
\affiliation{The Institute of Technological Science, Wuhan University, Wuhan 430072, China}

\author{Fang Lv}
\affiliation{Key Laboratory of Artificial Micro- and Nano-Structures of Ministry of Education, School of Physics and Technology, Wuhan University, Wuhan 430072, China}

\author{Yue Hou}
\affiliation{The Institute of Technological Science, Wuhan University, Wuhan 430072, China}

\author{Bin Zhou}\email{binzhou@hubu.edu.cn}
\affiliation{Department of Physics, Hubei University, Wuhan 430062, China}
\affiliation{Key Laboratory of Intelligent Sensing System and Security of Ministry of Education, Hubei University, Wuhan 430062, China}

\author{Ziyu Wang}\email{zywang@whu.edu.cn}
\affiliation{Key Laboratory of Artificial Micro- and Nano-Structures of Ministry of Education, School of Physics and Technology, Wuhan University, Wuhan 430072, China}
\affiliation{The Institute of Technological Science, Wuhan University, Wuhan 430072, China}

\author{Rui Xiong}\email{xiongrui@whu.edu.cn}
\affiliation{Key Laboratory of Artificial Micro- and Nano-Structures of Ministry of Education, School of Physics and Technology, Wuhan University, Wuhan 430072, China}

\begin{abstract}
Recently, the structural disorder-induced topological phase transitions in periodic systems have attracted much attention. However, in aperiodic systems such as quasicrystalline systems, the interplay between structural disorder and band topology is still unclear. In this work, we investigate the effects of structural disorder on a quantum spin Hall insulator phase and a higher-order topological phase in a two-dimensional Amman-Beenker tiling quasicrystalline lattice, respectively. We demonstrate that the structural disorder can induce a topological phase transition from a quasicrystalline normal insulator phase to an amorphous quantum spin Hall insulator phase, which is confirmed by bulk gap closing and reopening, robust edge states, quantized spin Bott index and conductance. Furthermore, the structural disorder-induced higher-order topological phase transition from a quasicrystalline normal insulator phase to an amorphous higher-order topological phase characterized by quantized quadrupole moment and topological corner states is also found.  More strikingly, the disorder-induced higher-order topological insulator with eight corner states represents a distinctive topological state that eludes realization in conventional crystalline systems. Our work extends the study of the interplay between disorder effects and topologies to quasicrystalline and amorphous systems.

\end{abstract}

\maketitle

\section{Introduction}

Numerous factors can instigate topological phase transitions (TPTs) \cite{RevModPhys.80.167,RevModPhys.88.021004,RevModPhys.82.3045,RevModPhys.83.1057,PhysRevLett.122.237601,
PhysRevB.100.195432}, and notably, disorder-induced TPTs have attracted significant attention. This stems from the inherent presence of disorder to varying degrees in real materials. The Anderson-type on-site disorder, which is typically utilized to modulate the topological properties of electronic wavefunctions by altering the electronic chemical potential, leading to the theoretical prediction of the topological Anderson insulator as well as higher-order topological Anderson insulator in various condensed matter systems \cite{PhysRevLett.102.136806,PhysRevLett.103.196805,PhysRevLett.105.115501,PhysRevB.84.035110,PhysRevB.92.085410,PhysRevLett.116.066401,
PhysRevB.100.054108,PhysRevB.96.205304,PhysRevLett.105.216601,PhysRevLett.113.046802,PhysRevLett.115.246603,PhysRevB.95.245305,
PhysRevB.97.235109,PhysRevB.98.235159,PhysRevB.97.024204,PhysRevB.93.125133,qin2016disorder,PhysRevB.98.134507,PhysRevB.100.205302,
PhysRevA.101.063612,PhysRevLett.125.166801,PhysRevB.103.085408,PhysRevB.104.245302,2309.03688}, is widely employed to induce TPTs. However, due to the stringent requirements of experimental conditions, the topological Anderson insulator and higher-order topological Anderson insulator have only been achieved in one-dimensional ultracold atomic wires \cite{meier2018observation} and simulated experiments based on photonic crystals \cite{stutzer2018photonic,PhysRevLett.125.133603,PhysRevLett.129.043902}, phononic crystals \cite{AdvMater.202001034} and electric circuit setup \cite{PhysRevLett.126.146802}.

Recently, the TPT induced by structural disorder has attracted widespread attention. For instance, Li et.al have proposed a TPT induced by the structural disorder in one-dimensional amorphous Rydberg atom chain \cite{PhysRevLett.127.263004}. Later, the structural disorder induced a second-order topological insulator phase with topological hinge states in three-dimensional cubical lattice \cite{PhysRevLett.126.206404} and quantum spin Hall insulator phase with topological edge states in two-dimensional trigonal lattice \cite{PhysRevLett.128.056401} have been proposed. The origin of structural disorder-induced TPTs is the renormalization of spectral gaps associated with single-particle energy levels, resulting in potential changes in the energy levels and spectral gaps of electrons \cite{PhysRevLett.128.056401}. Therefore, the physical mechanisms of TPTs induced by structural disorder and Anderson-type on-site disorder are fundamentally different, with the latter arising from many-body effects, as the fluctuation of on-site energies is intricately linked to electron-electron interactions \cite{PhysRevLett.128.056401}. From a theoretical perspective, one direct approach to introducing structural disorder is to add a random displacement of varying magnitudes to each lattice point, causing the original lattice points to deviate from their initial positions, and accompanied by the amorphization of the material \cite{PhysRevLett.128.056401,PhysRevLett.126.206404}. The attractiveness of TPTs induced by structural disorder can be ascribed for two main reasons. Firstly, amorphous structures are ubiquitous in the natural world \cite{zallen2008physics}. Secondly, nearly all materials can be prepared into amorphous phases through various amorphization techniques \cite{cohen1964metastability,10.1063/1.1735777,jones1984status,Inoue_1988,chen1970formation,SCHULTZ198815,graves1994plasma,
toh2020synthesis}. Therefore, combining TPTs with structural disorder allows for observation in experiments using a wide range of selectable materials.

The quasicrystalline systems present new opportunities for exploring novel topological states associated with high rotational symmetry which is possing forbidden in crystals such as the fivefold \cite{PhysRevLett.129.056403}, eightfold \cite{PhysRevLett.124.036803}, and twelvefold \cite{PhysRevB.102.241102} rotational symmetries. As so far, the TPTs induced by Anderson-type on-site disorder in quasicrystalline systems have been extensively studied \cite{PhysRevB.100.115311,PhysRevA.105.063327,PhysRevA.106.L051301,PhysRevB.104.245302,PhysRevB.103.085307,PhysRevB.104.155304}.
Furthermore, several researches have investigated TPTs induced by structural disorder in the context of a periodic lattice framework \cite{PhysRevLett.128.056401,PhysRevLett.126.206404,2312.08779}. However, it is not yet clear whether structural disorder can induce TPTs in quasiperiodic lattice systems.

In this paper, we investigate the structural disorder-induced TPTs in an Ammann-Beenker tiling octagonal quasicrystal. In the case of the TPT from a quasicrystalline normal insulator phase to an amorphous quantum spin Hall insulator phase, by calculating the real-space spin Bott index \cite{PhysRevLett.121.126401,PhysRevB.98.125130}, two-terminal conductance \cite{Landauer_1970,B_ttiker_1988,Fisher_1981} and the probability density of the in-gap eigenstates, we find that the topologically trivial phase converts to a topologically non-trivial phase characterized by a nonzero spin Bott index ($B_{s}=1$) with a quantized conductance plateau ($G={2e^2}/{h}$) in a certain area of disorder strength. Furthermore, by calculating the real-space quadrupole moment \cite{PhysRevB.103.085408,PhysRevB.100.245135,PhysRevB.101.195309,PhysRevB.100.245134,PhysRevResearch.2.012067,PhysRevLett.125.166801} and the probability distribution of in-gap states, we have also identified a structural disorder-induced TPT from an initial quasicrystalline topologically trivial phase to an amorphous higher-order topological phase, occurring at a certain area of disorder strength with four localized gapless corner states characterized by quantized quadrupole moment ($q_{xy}=0.5$). Even more remarkable is the discovery of a structural disorder-induced higher-order topological insulator phase, characterized by eight localized gapless corner states confined within an octagonal boundary, representing a distinctive topological state that eludes realization in conventional crystalline systems.

The rest of the paper is organized as follows. We introduce a quantum spin Hall insulator model and a higher-order insulator model with structural disorder in two-dimensional quasicrystalline lattice and give the details of numerical methods in Sec.~\ref{Models}. Then, we provide numerical results for studying the TPTs of the two models in Sec.~\ref{Model1} and Sec.~\ref{Model2}, respectively. Finally, we summarize our conclusions in Sec.~\ref{Conclusion}.

\section{Models and Method}
\label{Models}
\begin{figure}[tpb]
   	\includegraphics[width=8cm]{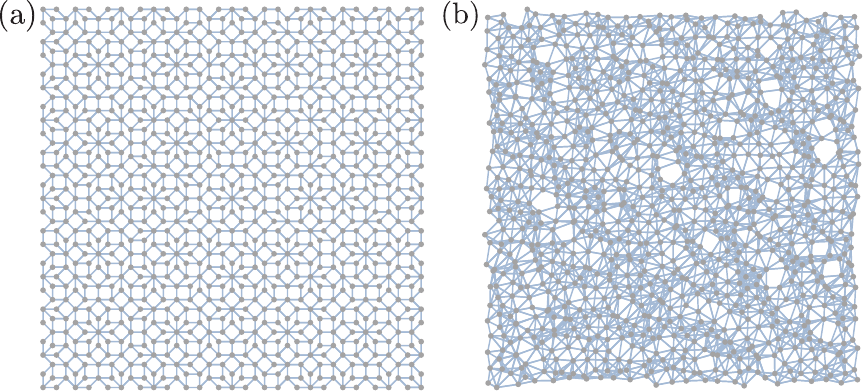} \caption{(a) Schematic diagram of the Ammann-Beenker tiling quasicrystal containing 1005 sites. The first three nearest-neighbor intercell bonds correspond to the short diagonal of the rhombus tile, the edge of square and rhombus tile, and the diagonal of square tile, respectively. The distance ratio of the three bonds is $r_{0}:r_{1}:r_{2}=2\sin \frac{\pi }{8}:1:2\sin \frac{\pi }{4}$. (b) The Ammann-Beenker tiling quasicrystal transforms into an amorphous structure in the presence of structural disorder, where the strength of disorder is $\sigma=0.2$. We set the hopping radius as $R=1.7$, ensuring the first three nearest-neighbor hopping in the quasicrystalline structure. To clearly depict the quasicrystalline structure, only the second-order nearest-neighbor bonds are illustrated in (a).}%
\label{fig1}
\end{figure}

We start with a tight-binding model of a quantum spin Hall insulator in an Ammann-Beenker tiling quasicrystalline lattice \cite{grunbaum1987tilings,c641bbfabe714fefbd7c37e571cb29fa,Duneau_1989,kramer2003coverings} with square boundary condition as shown in Fig.~\ref{fig1}(a). The quasicrystal structure is formed by the aperiodic arrangement of squares and rhombi with equal side lengths in a two-dimensional plane. We consider the first three nearest-neighbor hopping, i.e. the short diagonal of the rhombus, the edge of the rhombus and square, and the diagonal of the square. The model Hamiltonian is given by
\begin{eqnarray}
H &=&-\sum_{m\neq n}\frac{l(r_{mn})}{2}c_{m}^{\dagger }[it_{1}(s_{3}\tau
_{1}\cos \psi _{mn}+s_{0}\tau _{2}\sin \psi _{mn}) \notag \\
&&+t_{2}s_{0}\tau _{3}]c_{n}+\sum_{m}(M+2t_{2})c_{m}^{\dagger }s_{0}\tau _{3}c_{m},
\label{H}
\end{eqnarray}
where $c_{m}^{\dag }=(c_{m\alpha \uparrow }^{\dag },c_{m\alpha \downarrow }^{\dag
},c_{m\beta \uparrow }^{\dag },c_{m\beta \downarrow }^{\dag })$ represents the creation operator of an electron on a site $m$. $m$ and $n$ denote lattice sites running from $1$ to $N$, and $N$ is the total number of lattice sites. In each site, $\alpha$ ($\beta$) is the index of orbitals and $\uparrow$ ($\downarrow$) represents the spin direction. $s_{1,2,3}$ and $\tau_{1,2,3}$ are the Pauli matrices acting the spin and orbital degree of freedom, respectively. $s_{0}$ and $\tau_{0}$ are the $2\times 2$ identity matrices. $t_{1,2}$ are the hopping strength and $M$ is the Dirac mass. $\psi_{mn}$ is the polar angle of bond between site $m$ and $n$ with respect to the horizontal direction. $l(r_{mn})=e^{1-r_{mn}/\lambda }$ is the spatial decay factor of hopping amplitudes with the decay length $\lambda $, where $r_{mn}=|\mathbf{r}_{m}-\mathbf{r}_{n}|$ is the distance from the site $m$ to site $n$.  The model Hamiltonian preserves time-reversal symmetry $T$, particle-hole symmetry $P$ and chiral symmetry $S$, therefore it belongs to the class DIII \cite{PhysRevB.55.1142,PhysRevB.78.195125}. Here, the symmetry operators are $T=is_{2}\tau_{0}K$, $P=s_{3}\tau_{1}K$ and $S=TP$, respectively, where $K$ is the complex conjugate operator. Furthermore, when the Hamiltonian (\ref{H}) is projected onto a square lattice with exclusive consideration of nearest-neighbor hopping, the model will evolve into a quantum spin Hall insulator based on HgTe quantum wells \cite{bernevig2006quantum}. Without loss of generality, the spatial decay length $\lambda$ and the side length of rhombus and square $r_{1}$ are fixed as $1$, and the energy unit is set as $t_{1}=t_{2}=1$.

In order to investigate TPTs in quasicrystalline lattice with varying degrees of structural disorder, we considered the atomic thermal fluctuations corresponding to a typical quenching process of a molten state. The specific approach involves adding a displacement $\textbf{r}$ at each lattice point in the quasicrystal structure with a randomly determined magnitude and direction, where the magnitude follows a Gaussian distribution \cite{PhysRevLett.128.056401,PhysRevB.108.L081110}. The random displacement can be written as
\begin{eqnarray}
D(\textbf{r}) &=&\frac{1}{2\pi\sigma^{2}}\rm{exp}\left(-\frac{\textit{r}^{2}}{2\sigma^{2}}\right),
\label{D}
\end{eqnarray}
where the distance standard deviation $\sigma$ is the intensity of structural disorder which is scaled in unit of the edge of the rhombus and square. In Fig.~\ref{fig1}(b), we plot the Ammann-Beenker tiling quasicrystal in the presence of structural disorder with the strength of disorder being $\sigma=0.2$, where the quasicrystalline structure has been converted to an amorphous structure.

To characterize the structural disorder-induce amorphous quantum spin Hall insulator phase, we adopt the spin Bott index \cite{PhysRevLett.121.126401,PhysRevB.98.125130} as well as the two-terminal conductance based on the recursive Green function method \cite{MacKinnon_1985,PhysRevB.72.235304}. The detailed steps of numerical calculation of the spin Bott index can be summed up as follows. First, one constructs the projector operator of the occupied states as $P=\sum_{i}^{N_{occ}}|\psi _{i}\rangle \langle \psi _{i}|$, where $\psi _{i}$ is the $i$th wave function of the Hamiltonian (\ref{H}) and $N_{occ}$ is the total number of occupied states. Second, one introduces another projector operator as $P_{z}=P\hat{\eta}_{z}P$, where $\hat{\eta}_{z}=\frac{\hbar }{2}s _{3}$ is the spin operator with the Pauli matrix $s _{3}$. The eigenvalues of $P_z$ are divided into two parts by zero energy, in which the number of positive and negative eigenvalues are both equal to $N_{occ}/2$. Then, a new projector operator can be constructed as $P_{\pm }=\sum_{i}^{N/2}\left\vert \phi _{i}^{\pm }\right\rangle \left\langle \phi _{i}^{\pm }\right\vert$. The projected position operators of the two spin sectors can be defined as
\begin{align}
&U_{\pm }=P_{\pm }e^{i2\pi X}P_{\pm }+(I-P_{\pm }),\\
&V_{\pm }=P_{\pm }e^{i2\pi Y}P_{\pm }+(I-P_{\pm }),
\label{UV}
\end{align}
where $X$ and $Y$ are two diagonal matrices, $X_{ii}=x_{i}/L_{x}$ and $Y_{ii}=y_{i}/L_{y}$ with $(x_{i},y_{i})$ being the coordinate of the $i$th lattice site, and $L_{x(y)}$ being the the size of the sample along the $x(y)$ direction. Finally, one can obtain the spin Bott index as
\begin{align}
B_{s}=\frac{1}{2}(B_{+}-B_{-}),
\label{Bott}
\end{align}
where $B_{\pm }=\frac{1}{2\pi }{\rm Im}\{{\rm Tr}[\ln (\tilde{V}_{\pm }\tilde{U}_{\pm }\tilde{V}_{\pm }^{\dag
}\tilde{U}_{\pm }^{\dag })]\}$
are the Bott indexes of spin up and down, respectively. The case with $B_{s}=0$ corresponds to the normal insulator phase, and $B_{s}=1$ corresponds to the quantum spin Hall insulator phase. It is noted that the calculation of $B_{s}$ is performed under the framework of periodic boundary conditions constructed using quasi-periodic approximation theory \cite{JPSJ.55.1420,PhysRevB.43.8879,refId0}.

Morover, according to Landauer-B\"uttiker-Fisher-Lee formula \cite{Landauer_1970,B_ttiker_1988,Fisher_1981}, the conductance can be written as
\begin{align}
G=\frac{e^2}{h}T(\mu ),
\label{Conductance}
\end{align}
where $T(\mu )= \text{Tr}[\Gamma _{L}(\mu )G^{r}(\mu )\Gamma _{R}(\mu )G^{a}(\mu )]$ is the transmission coefficient at energy $\mu$. $\Gamma _{L(R)}(\mu)=i(\Sigma_{L(R)}^{r}-\Sigma_{L(R)}^{a})$ is the left (right) linewidth with the left (right) lead retarded self-energy $\Sigma^{r} _{L(R)}$ and the left (right) lead advanced self-energy $\Sigma^{a} _{L(R)}$. $G^{r(a)}(\mu)$ is the retarded (advanced) Green's function of the device, and can be expressed as
\begin{align}
G^{r}(\mu )=[G^{a}(\mu )]^{\dag }=[\mu -H_{d}-\Sigma _{L}^{r}-\Sigma_{R}^{r}]^{-1},
\label{r-a-Conductance}
\end{align}
where $H_{d}$ is the device Hamiltonian. For a quantum spin Hall insulator phase, the two-terminal conductance is a quantized number $G={2e^2}/{h}$, and $G=0$ corresponds to a normal insulator phase.

It has been proposed that the Wilson mass term can destroy the time-reversal symmetry of the quantum spin Hall insulator which is described by Hamiltonian (\ref{H}), so that the original helical boundary state of the system opens the energy gap and evolves into a higher-order corner state \cite{PhysRevLett.124.036803}. The mass term can be written as
\begin{align}
H_{g}=-g\sum_{m\neq n}\frac{l(r_{mn})}{2}c_{m}^{\dagger }s_{1}\tau _{1}\cos (\xi \psi _{mn})c_{n}
\label{Hg}
\end{align}
where $g$ is the magnitude of the mass term and $\xi$ is the varying period of the mass term, and $\xi = 2$ (4) for square (octagonal) samples.

To describe higher-order topological phase in a aperiodic lattice, we will employ topological invariant real-space quadrupole moment \cite{PhysRevB.103.085408,PhysRevB.100.245135,PhysRevB.101.195309,PhysRevB.100.245134,PhysRevResearch.2.012067,PhysRevLett.125.166801}. The form of real-space quadrupole moment is
\begin{equation}
q_{xy}=\frac{1}{2\pi }{\rm{Im}} \log [\det (\Psi _{occ}^{\dagger }\hat{U}\Psi
_{occ})\sqrt{\det (\hat{U}^{\dagger })}],
 \label{qxy}
\end{equation}
where $\Psi _{occ}$ is the eigenvectors of occupied states. $\hat{U}\equiv \exp [i2\pi \hat{X}\hat{Y}/L^{2}]$ where $\hat{X}$ and $\hat{Y}$ are the position operator and $L$ represents the side length of the sample. In the case of $q_{xy}=0.5$, the system is a second-order topological phase with topological corner states. Besides, $q_{xy}=0$ indicates a trivial phase. Note that the subsequent calculations of $q_{xy}$ in this paper are based on periodic boundary conditions.

In addition, another suitable method for characterizing higher-order topological phase is to use the existence of corner states as a criterion for determination  \cite{PhysRevB.99.085406,PhysRevLett.126.146802,PhysRevB.106.125310,PhysRevB.107.125302,PhysRevB.104.245302}. In the subsequent computations related to disorder-induced higher-order topological phases, we determine the topology of the system by computing the real-space topological invariant $q_{xy}$ as well as examining the presence of corner states.

\section{structural disorder-induced first-order topological phase}
\label{Model1}
In this section, we focus on the stuctural disorder-induced first-order topological insulator phase in an Ammann-Beenker tiling quasicrystal with square boundary condition. We have confirmed that even in the presence of structural disorder, symmetries such as time reversal symmetry, particle-hole symmetry and chiral symmetry remain preserved.

\begin{figure}[tpb]
	\includegraphics[width=8.5cm]{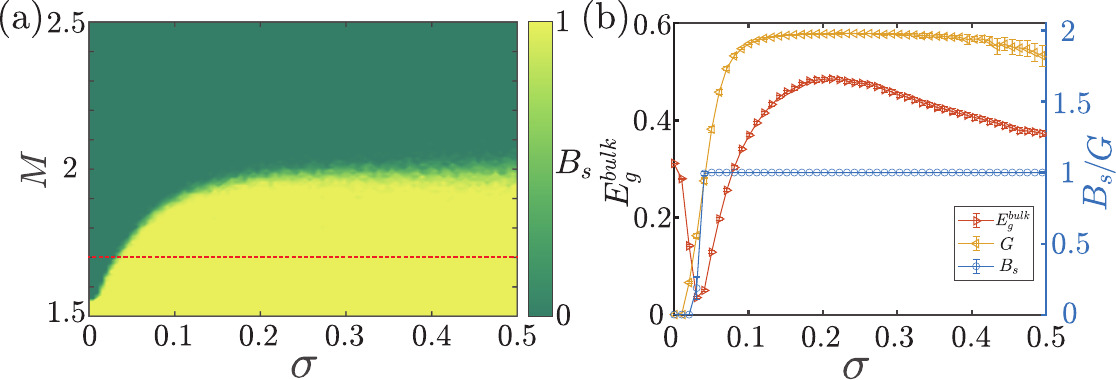} \caption{(a) Topological phase diagram of the Ammann-Beenker tiling quasicrystal in ($M, \sigma$) space obtained by calculating the real-space topological invariant spin Bott index $B_{s}$. The yellow color represents the topological insulator phase corresponding to $B_{s}=1$, and the green color represents the normal insulator phase corresponding to $B_{s}=0$. The red dashed line represents $M=1.7$. (b) Bulk energy gap $E^{bulk}_{g}$, spin Bott index $B_{s}$ and two-terminal conductance $G$ versus disorder strength $\sigma$ with $M=1.7$. The system with PBC marked by the orange line shows that the bulk energy gap has undergone the process of closing and reopening. The system contains $1005$ sites and $500$ disorder configurations are made.}%
\label{fig2}
\end{figure}

Figure~\ref{fig2}(a) shows the spin Bott index $B_{s}$ as a function of disorder strength $\sigma$ and Dirac mass $M$. The color map shows the magnitude of the spin Bott index. In the clean limit, i.e. $\sigma=0$, the quasicrystalline lattice hosts a normal insulator phase characterized by $B_{s}=0$ with the parameter $1.55<M<2$. However, as the strength of structural disorder increasing from $0$, the normal insulator phase converts to a topological insulator phase characterized by $B_{s}=1$. It is noted that the TPTs are limited to the region $1.55 < M < 2$. Additionally, we observed that the critical disorder strength $\sigma_{c}$ of the TPTs increases with the growth of $M$.

It is well known that the process of TPT is inevitably accompanied by the closing and reopening of the bulk energy gap, and the point where the gap closes represents the critical point of the phase transition. In general, the energy gap of a system with periodic boundary conditions is equal to the bulk energy gap. However, for aperiodic systems, it is necessary to employ quasiperiodic approximation theory to construct periodic boundary conditions. In Fig.~\ref{fig2}(b), we plot the bulk energy gap versus disorder strength marked by orange right triangles with $M=1.7$ marked by red dashed line in Fig.~\ref{fig2}(a). In the clean limit, the system hosts a normal insulator band gap with $E^{bulk}_{g}\approx 0.32$ . With the increasing of disorder strength, the bulk energy gap monotonically decreases until the critical point $\sigma\approx 0.04$, beyond which the bulk energy gap gradually increases. The evolution of the bulk energy gap conforms to the process of gap closure and reopening, indicating that the system undergoes a TPT during the process of disorder enhancement. However, we must point out that, in our calculations, the minimum value of the bulk energy gap is not strictly equal to zero. A reasonable explanation for this issue is the presence of finite-size effects \cite{PhysRevLett.101.246807}. When a sufficiently large sample size is chosen, the size effects can be significantly mitigated [see Fig.~\ref{fig7} in the Appendix].

We also compute a line graph depicting the evolution of the real-space topological invariant spin Bott index varying disorder strength [marked by blue circles in Fig.~\ref{fig2}(b)], which matches well with the evolution process of the bulk energy gap. It is found that the spin Bott index jumps from $0$ to $1$ at $\sigma\approx 0.04$ and then remains stable thereafter, with a platform without any fluctuations. Thus, it is further indicated that the topological phase transition induced by structural disorder has occurred. Furthermore, we map a line graph showing the variation of two-terminal conductance with changing disorder strength [marked by yellow left triangles in Fig.~\ref{fig2}(b)]. As the disorder strength reaches the phase transition critical point, the conductance gradually increases from $0$ and approaches ${2e^2}/{h}$. Subsequently, a typical quantized conductance plateau at $G={2e^2}/{h}$ emerges, providing further evidence for the structural disorder-induced phase transition from a normal insulator phase to a quantum spin Hall insulator phase.

\begin{figure}[tp]
	\includegraphics[width=7.5cm]{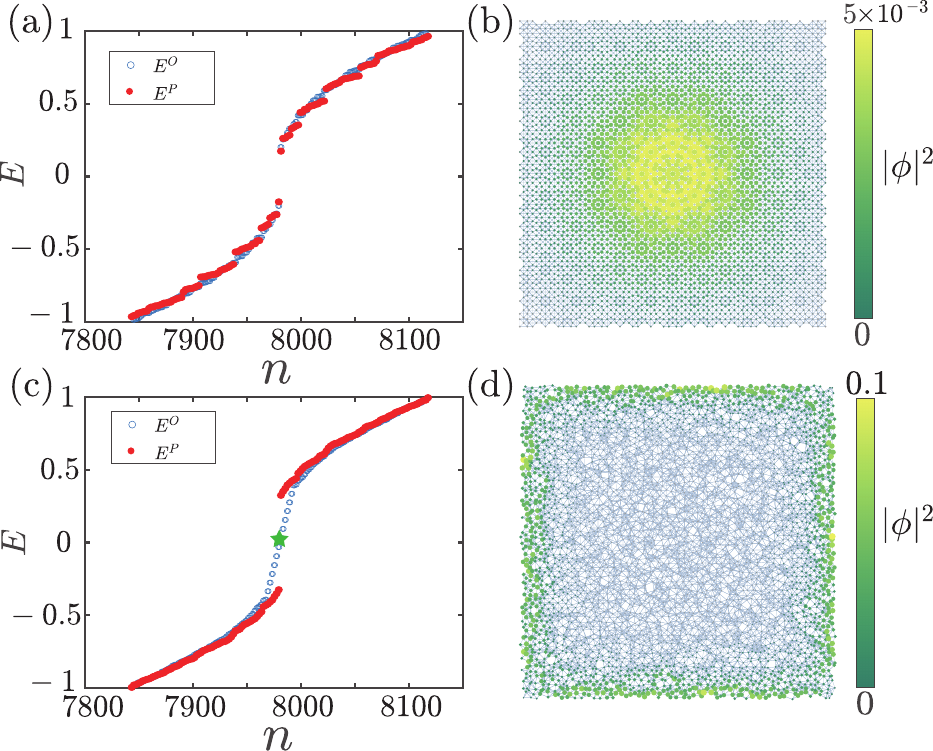} \caption{(a) Energy spectrum of the normal insulator with open boundary condition (marked by blue circles) and periodic boundary condition (marked by red circles), corresponding to the situation of $\sigma=0$ in Fig.~\ref{fig2}(b). (b) The probability density of the four eigenstates which are the nearest to zero energy in (a). (c) Energy spectrum of the topological insulator corresponding to the situation of $\sigma=0.18$ in Fig.~\ref{fig2}(b). (d) The probability density of the edge states marked by green star. To mitigate finite-size effect, the system size is set to contain $4061$ sites, and $M=1.7$. The colorbar represents the magnitude of the probability of the wavefunction $|\phi|^{2}$.}%
\label{fig3}
\end{figure}

For a more intuitive presentation of the aforementioned computational results, we plot the energy spectrum and wave functions distribution of the normal insulator phase as well as the disorder-induced topological insulator phase, shown in Fig~\ref{fig3}. When $\sigma=0$, the system hosts a normal insulator phase with large energy gap in both open boundary conditions (marked by blue circles) and periodic boundary conditions (marked by red dots), as shown in Fig~\ref{fig3}(a). The corresponding values of spin Bott index and conductance are equal to zero. In the scenario, the wave functions are localized in the bulk. However, when the disorder strength is set as $\sigma=0.18$, a series of in-gap states within the bulk energy gap have emerged [as shown in Fig~\ref{fig3}(c)], localized at the boundaries of the sample [see Fig~\ref{fig3}(d)], indicating that the system is now in a topological insulator phase. This computational result agrees well with the calculated values of spin Bott index and two-terminal conductance. Therefore, the TPT from a normal insulator phase to a topological insulator phase induced by structural disorder is clearly confirmed.

\section{structural disorder-induced higher-order topological phase}
\label{Model2}

In this section, we concentrate on the structural disorder-induced higher-order topological phase in an Ammann-Beenker tiling quasicrystal with square and octagon boundary condition. All calculations are based on the Hamiltonian $\mathcal{H}=H+H_{g}$. The introduction of the Willson mass term $H_{g}$ breaks the time-reversal symmetry and the chiral symmetry of the system, while the particle-hole symmetry is still preserved. Furthermore, we have also confirmed that structural disorder does not lead to the breaking of particle-hole symmetry. Therefore, the real-space quadrupole moment can be quantized, as it is protected by particle-hole symmetry \cite{PhysRevLett.125.166801}.

\begin{figure}[tp]
	\includegraphics[width=8.5cm]{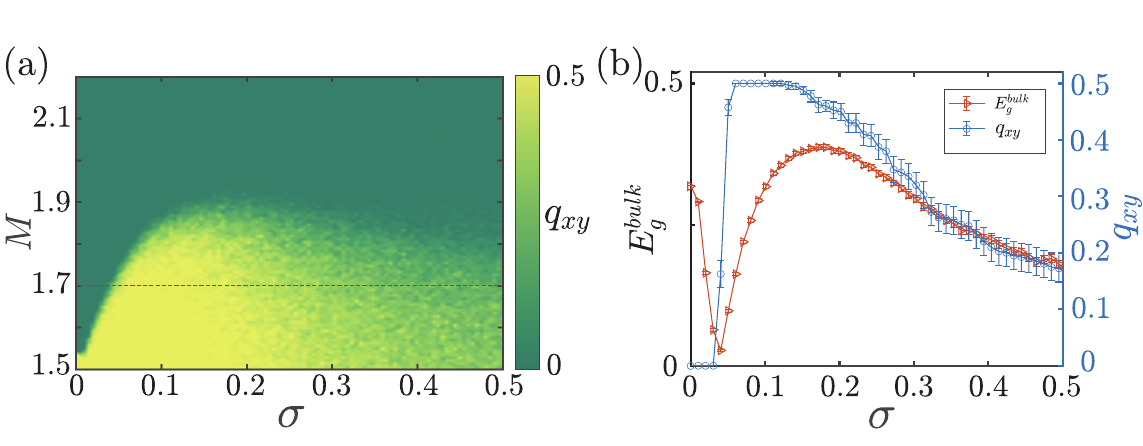} \caption{(a) Topological phase diagram of the Ammann-Beenker tiling quasicrystal in ($M, \sigma$) space obtained by calculating the real-space topological invariant quadrupole moment $q_{xy}$. The yellow color represents the higher-order topological insulator phase corresponding to $q_{xy}=0.5$, and the green color represents the normal insulator phase corresponding to $q_{xy}=0$. The red dashed line represents $M=1.7$. (b) Bulk energy gap $E^{bulk}_{g}$ marked by orange righet triangles and quadrupole moment $q_{xy}$ marked by blue circles versus disorder strength $\sigma$ with $M=1.7$. The system contains 1005 sites and 500 disorder configurations are made.}%
\label{fig4}
\end{figure}

We map out the phase diagram in $(M,\sigma)$ space for the case of square boundary condition in Fig.~\ref{fig4}(a), where $g=1$. The color map represents the magnitude of the real-space quadrupole moment $q_{xy}$. It is found that the system is in a topologically trivial phase with $q_{xy}=0$ in a region where $1.55<M<2.2$ in the clean limit. With the increase of the disorder strength, the higher-order topological phase occurs with $q_{xy}$ going from $0$ to $0.5$ in the region of $1.55<M<1.9$. The values of critical points increase monotonically with the increase of $M$. The structural disorder-induced higher-order insulator phases remain stable when the disorder strength is less than $0.2$, beyond which the higher-order topological phases gradually break down.

In Fig.~\ref{fig4}(b), we plot the real-space quadrupole moment $q_{xy}$ versus disorder strength $\sigma$ as well as the bulk energy gap with $M=1.7$ by red dashed line in Fig.~\ref{fig4}(a). The system is a normal insulator phase in the clean limit characterized by $q_{xy}=0$ and a bulk energy gap $E^{bulk}_{g}\approx 0.32$. With the increase of $\sigma$, in the region $0.045<\sigma<0.18$, a remarkable plateau of quantized $q_{xy}=0.5$ appear, which indicates a second-order topological phase induced by structural disorder. Meanwhile, the bulk energy gap monotonically decreases until the critical point $\sigma \approx 0.045$, beyond which the bulk energy gap gradually increases. The closure and subsequent reopening of the bulk energy gap further demonstrate the structural disorder-induced topological phase. However, as the disorder strength continues to increase, the quantized quadrupole moment platform gradually disappears, accompanied by a gradual reduction of bulk energy gap.

\begin{figure}[tp]
	\includegraphics[width=7.5cm]{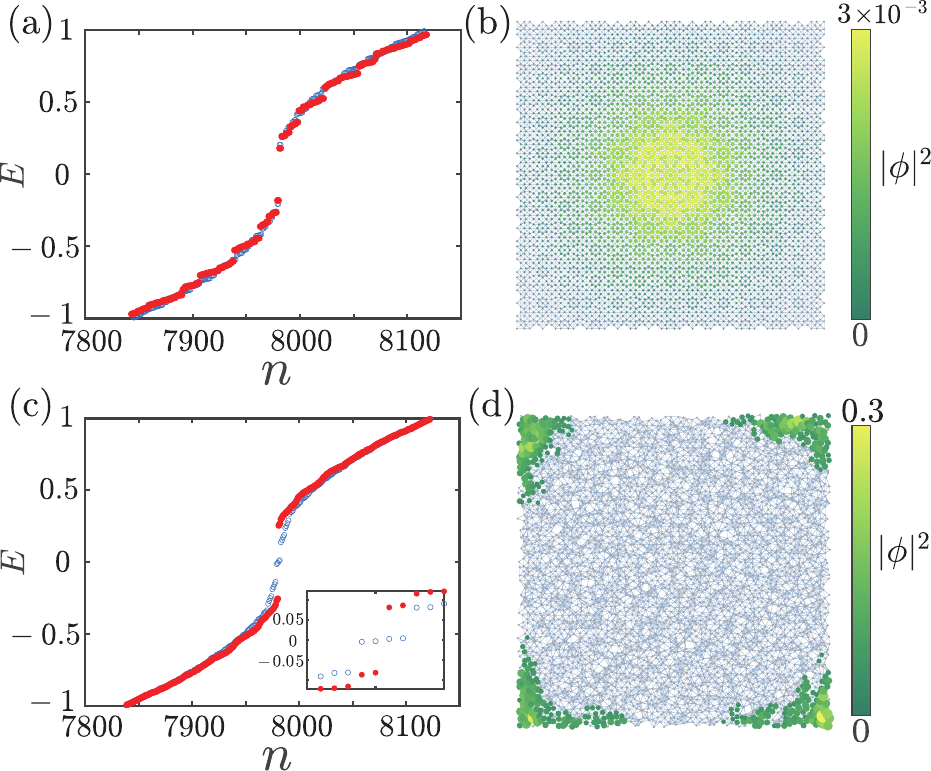} \caption{(a) Energy spectrum of the normal insulator with open boundary condition (marked by blue circles) and periodic boundary condition (marked by red dots), corresponding to the situation of $\sigma=0$ in Fig.~\ref{fig4}(b). (b) The probability density of the four eigenstates which are the nearest to zero energy in (a). (c) Energy spectrum of the topological insulator corresponding to the situation of $\sigma=0.15$ in Fig.~\ref{fig4}(b). (d) The probability density of the four eigenstates which are the nearest to zero energy . The system size is set to contain $4061$ sites and $g=1$, $M=1.7$. The colorbar represents the magnitude of the probability of the wavefunction $|\phi|^{2}$.}%
\label{fig5}
\end{figure}

To verify  the aforementioned computational results of quadrupole moment and bulk energy gap, we plot the eigenspectrum and probability density of the four eigenstates which are the nearest to zero energy in Fig.~\ref{fig5}. It is shown that the system hosts a normal insulator phase with large energy gap in both open boundary conditions and periodic boundary conditions, and the four eigenstates which are the nearest to zero energy are local in the bulk, as shown in Fig.~\ref{fig5}(a). However, when the disorder strength is set as $\sigma=0.15$, four in-gap states appear, localized at the four corners of the sample [see Figs.~\ref{fig5}(c) and (d)]. These corner states are the powerful proof of structural disorder-induced higher-order topological insulator. It is noted that the four in-gap states are not strictly degenerate to zero, which is owing to the finite size effect. For a sample of finite size, the four corner states tend to overlap. In the inner illustration of Fig.~\ref{fig5}(c), we plot the eigenspectrum of the open lattice contains $16437$ sites with $\sigma=0.15$. It is found that the energy values of the four in-gap states are approximately equal to zero energy.

\begin{figure}[tp]
	\includegraphics[width=7.5cm]{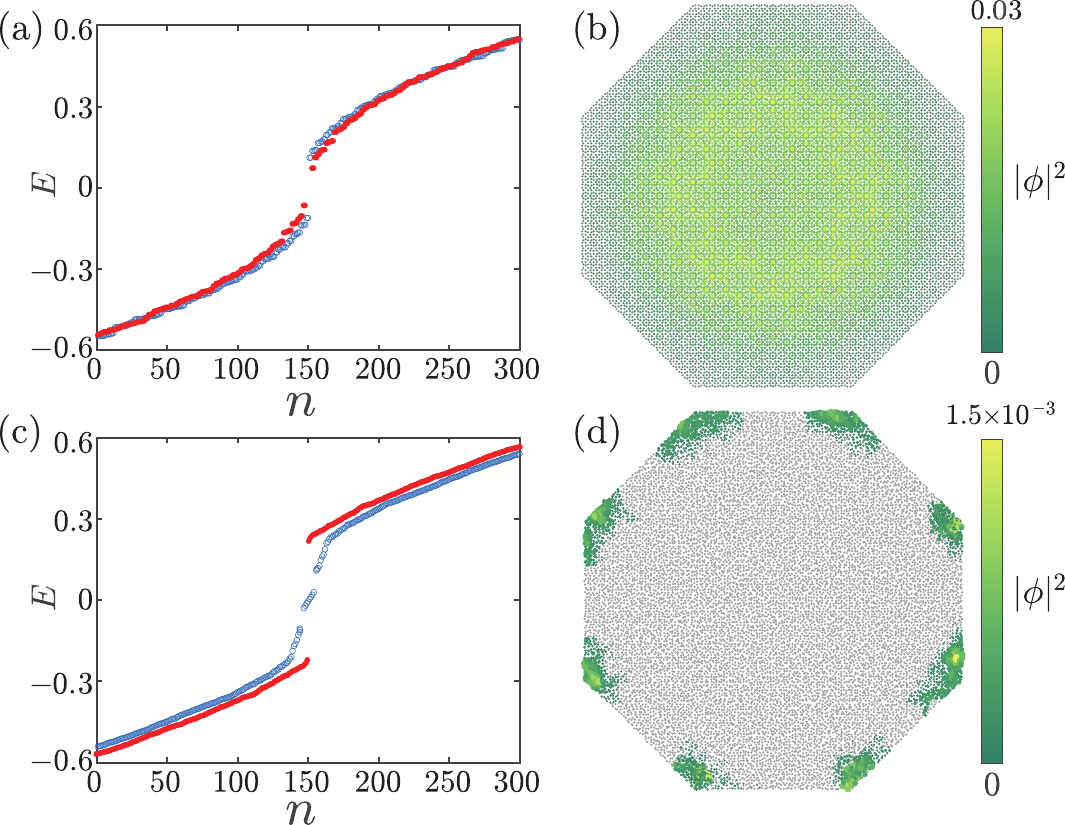} \caption{(a) Energy spectrum of the normal insulator with open boundary conditions (marked by blue circles) and periodic boundary conditions (marked by red dots), where $\sigma=0$. (b) The probability density of the eight eigenstates which are the nearest to zero energy in (a). (c) Energy spectrum of the higher-order topological insulator phase in open (periodic) boundary conditions, where $\sigma=0.13$. (d) The probability density of the eight eigenstates which are the nearest to zero energy . The system size is set to contain $13289$ sites and $g=1.4$, $M=1.55$. The colorbar represents the magnitude of the probability of the wavefunction $|\phi|^{2}$.}%
\label{fig6}
\end{figure}

In addition, we calculate structural disorder-induced higher-order topological phase in Ammann-Beenker tiling quasicrystal with octagonal boundary. As shown in Fig.~\ref{fig6}, we plot the eigenspectrum and probability density of the eight eigenstates which are the nearest to zero energy with different disorder strength. In the clean limit, i.e. $\sigma=0$, the system is in a topologically trivial phase characterized by a trivial energy gap both in open (marked by blue circles) and periodic boundary conditions (marked by red dots) with eigenstates localized in the bulkn shown in Fig.~\ref{fig6}(b). However, when the structural disorder strength is set as $\sigma=0.13$, a second-order topological phase arises with eight topological corner states respectively localized at the eight corners of the sample, as shown in Fig.~\ref{fig6}(d). We believe that higher-order topological insulator with eight corner states is a novel phase belongs exclusively to the quasicrystalline structure. The quadrupole moment $q_{xy}$ is only available to measure the bulk quadrupole topology with square boundary condition. A higher-order topological insulator with more than four zero energy modes does not belong to the class of topological quadrupole insulators. Thus, it does not work for octagonal samples. It is noted that a generalized quadrupole moment has been proposed to characterize the higher-order phase in a regular octagon \cite{SciPostPhys.15.5.193}, and such a method is used to characterize the Weyl semimetal phase in a three-dimensional quasicrystal \cite{2307.14974}. Another appropriate way to characterize the higher-order topological phase is to adopt the existence of the corner states as a working definition \cite{PhysRevB.99.085406,PhysRevLett.126.146802,PhysRevB.106.125310,PhysRevB.107.125302,PhysRevB.104.245302}. Thus,the eight corner states, induced by structural disorder in Fig.~\ref{fig6}(d), are strong evidence for the emergence of the higher-order topological insulator.

Additionally, we would like to point out that, despite employing a relatively large lattice model in the calculation of the octagonal boundary condition, the energy eigenvalues of corner states still do not degenerate to zero. We attribute it to the easier overlap of two corner states along each edge of the octagon (compared to a square), which is inherently the finite-size effect. We speculate that under thermodynamic limit conditions, the eight corner states will exhibit better localization, and concurrently, their energy eigenvalues will tend to be degenerate. In addition, we suspect that the structural disorder-induced higher-order topological insulator with eight corner states arise from the rotation symmetry of the quasicrystal. Despite the local disruption of the eightfold rotation symmetry due to structural disorder, the global symmetry is still maintained on a statistical average \cite{PhysRevB.106.195304,PhysRevLett.109.246605,PhysRevB.97.205110}. Thus, it can be anticipated that the symmetry, including fourfold and eightfold rotation symmetry, disrupted by structural disorder, is statistically restored through ensemble averaging \cite{SciPostPhys.15.5.193}, and the higher-order topological insulator phase is protected by the average global symmetry.

\begin{figure}[htb]
	\includegraphics[width=8.5cm]{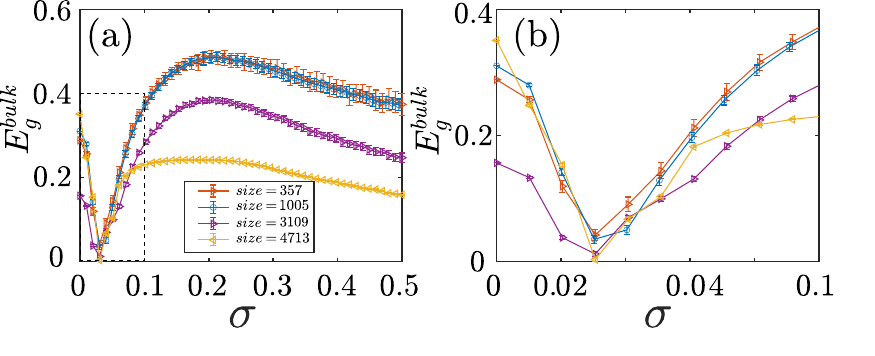} \caption{(a) The bulk energy gap as a function of disorder strength $\sigma$ with different system sizes. (b) An enlarged view corresponding to the dashed area in (a). An average of $400$ random configurations are taken. The other parameters are the same as in Fig. 2(b).}%
\label{fig7}
\end{figure}
\section{Conclusion}
In this work, we investigate the structural disorder-induced first-order topological phase and higher-order topological phase in an Ammann-Beenker tiling quasicrystal. The quasicrystalline quantum spin Hall insulator is considered to be the fundamental model. We introduce structural disorder into the normal insulator phase which is obtained by adjusting parameters. Based on calculating the real-space spin Bott index and two-terminal conductance, we demonstrate that structural disorder-induced first-order topological insulator phase characterized by quantized spin Bott index and quantized conductance has been found. Additionally, the states that emerge within the bulk energy gap localized at the four edges of the sample further validated our computational results. Furthermore, we introduce the Wilson mass term to the quasicrystalline quantum spin Hall insulator to obtain a higher-order topological insulator model, and an initial state with a normal insulator phase is achieved by modifying system parameters. Based on calculating the real-space quadrupole moment, we demonstrate that structural disorder-induced higher-order topological insulator phase characterized by quantized quadrupole moment has been found. The corner states localized at the vertices of sample serve as compelling evidence for the existence of higher-order topological insulators.
\label{Conclusion}

\section*{Acknowledgments}

This work was supported by the project from the NSFC (Grants No. 12122408) and the National Key R\&D Program (2023YFB4603800). B.Z. was supported by the NSFC (Grant No. 12074107) and the program of outstanding young and middle-aged scientific and technological innovation team of colleges and universities in Hubei Province (Grant No. T2020001), and the innovation group project of the Natural Science Foundation of Hubei Province of China (Grant No. 2022CFA012). T.P. was supported by the Doctoral Research Start-Up Fund of Hubei University of Automotive Technology (Grant No. BK202216). C.-B.H. was supported by the NSFC (Grant No. 12304539). Z.-R.L. was supported by the Postdoctoral Fellowship Program of CPSF (under Grant No. GZC20230751) and the Postdoctoral Innovation Research Program in Hubei Province (under Grant No. 351342).

\section*{Appendix: Finite-size analysis }

In this Appendix, we plot the bulk energy gap versus the size of the system. It has been observed that the minimum value of the bulk energy gap decreases with increasing system size, as shown in Fig.~\ref{fig7}. It can be inferred that under the conditions of the thermodynamic limit, the bulk energy gap will be closed strictly at the critical point of phase transition.

\bibliographystyle{apsrev4-1-etal-title_6authors}

\end{document}